\documentstyle[12pt,titlepage]{article}

\font\grande=cmr9.5 scaled \magstep4
\font\medio=cmr9.5 scaled \magstep2
\outer\def\beginsection#1\par{\medbreak\bigskip
      \message{#1}\leftline{\bf#1}\nobreak\medskip
\vskip-\parskip
      \noindent}

\def\laq{\raise 0.4ex\hbox{$<$}\kern -0.8em\lower 0.62
ex\hbox{$\sim$}}

\def\gaq{\raise 0.4ex\hbox{$>$}\kern -0.7em\lower 0.62
ex\hbox{$\sim$}}

\begin{document}
\bibliographystyle {unsrt}

\titlepage

\vspace{15mm}
\begin{center}
{\grande Backgrounds of squeezed relic photons }\\
\vspace{4mm}
{\grande and their spatial correlations}\\
\vspace{15mm}

 Massimo Giovannini 
\footnote{Electronic address: giovan@cosmos2.phy.tufts.edu} \\
\vspace{6mm}

{\sl  Institute of Cosmology, Department of Physics and Astronomy, \\
Tufts University, Medford, Massachusetts 02155}\\

\end{center}

\vskip 2cm
\centerline{\medio  Abstract}

\noindent
We discuss the production of multi-photons 
squeezed states induced by the time  variation of the (Abelian)
gauge coupling constant in a string cosmological context. 
Within a fully quantum mechanical approach
we solve the time evolution of the mean number 
of produced photons in terms of the squeezing parameters and in terms of 
the gauge coupling. We compute the first (amplitude interference) 
and second order (intensity interference) 
correlation functions of the magnetic part of the photon background. 
The photons produced 
thanks to the variation of the dilaton coupling are strongly 
 bunched for the realistic case where the growth of the 
dilaton coupling is required to explain the presence 
of large scale magnetic fields and, possibly of a Faraday rotation of the
Cosmic Microwave Background.

\vspace{5mm}

\vfill
\newpage

The squeezed states formalism has been successfully
applied to the analysis of tensor,  scalar \cite{gr} 
and rotational \cite{gr2} 
fluctuations of the metric 
by Grishchuk and collaborators. 
In this paper we want to address 
the possible application of the squeezed states formalism 
to the case of relic photons which, to the best of our knowledge, 
has not received particular attention. In the case 
of relic gravitons and relic phonons the analogy with
quantum optics is certainly very inspiring. In the 
case of relic photons the analogy is even closer since 
the time variation of the dilaton coupling plays directly the 
r\^ole of the laser ``pump'' which is employed in order 
to produce experimentally observable squeezed states \cite{sq}. 

In a general relativistic context the evolution 
 equations of the (Abelian)
gauge field strength is invariant under conformal (Weyl)
rescaling of the metric tensor.  
If we are in a conformally flat 
background of Friedmann-Robertson-Walker (FRW) 
type (written in conformal time $\eta$) 
\begin{equation}
ds^2 = G_{\mu\nu} dx^{\mu} d x^{\nu} = a^2(\eta) [ d\eta^2 - d\vec{x}^2],
\label{metric}
\end{equation}
the evolution of the Abelian gauge field strength 
 implies that, in vacuum, the determinant 
of the FRW metric can be always reshuffled by appropriately 
rescaling the gauge fields so that the (rescaled)
 magnetic and electric 
field amplitudes will always  obey, in the curved background  
of Eq. (\ref{metric}) the same Maxwell's equations they would 
obey in flat space-time. Consequently, 
in general relativity Abelian gauge fields cannot be 
amplified from their vacuum fluctuations \cite{gr1,par}. 
On the contrary, tensor fluctuations of the metric are 
amplified in an isotropic FRW space since their evolution 
equation is not invariant under Weyl rescaling of the 
background metric \cite{gr1}.

A  peculiar feature of cosmological models 
inspired by the low energy string effective action is that
the gauge coupling is not really a constant but it evolves 
in time \cite{ven}. 
Therefore, in string cosmology, photons \cite{1} (as well as gravitons 
\cite{9}) can be produced
thanks to the time evolution of the gauge coupling from the 
electromagnetic vacuum fluctuations. 
A similar amplification effect exists 
for the case of non-Abelian gauge fields which are, however, 
screened at high temperature. 
The large (but finite) value of the conductivity in the early Universe 
implies that the Abelian
magnetic component survives only for momenta which are 
much shorter than the magnetic diffusivity scale at each epoch \cite{kra,zel}.
Plasma effects connected with the evolution of large scale magnetic fields
have been explored with particular attention to the 
electroweak scale \cite{2}.
The effective 
action of a generic Abelian gauge field in four space-time 
dimensions reads 
\begin{equation}
S = - \frac{1}{4} \int d^4 x \sqrt{- G} \frac{1}{g^2} 
F_{\alpha\beta} ~F^{\alpha\beta},
\label{action}
\end{equation}
where $F_{\alpha\beta} = \nabla_{[\alpha} A_{\beta]}$ is the
 Maxwell field strength and $\nabla_{\alpha}$ is the covariant 
derivative with respect to the string frame metric $G_{\mu\nu}$. 
In Eq. (\ref{action}) $g= \exp{(\phi/2)} $ is the (four dimensional)
 dilaton coupling. 

In the absence of a classical 
gauge field background, the perturbed effective Lagrangian density 
\begin{eqnarray}
&&{\cal L}(\vec{x}, \eta) = 
\frac{1}{2}\sum_{\alpha} \biggl[{{\cal A}'}_{\alpha}^2 +
 2 \frac{g'}{g} {\cal A}'_{\alpha} {\cal A}_{\alpha} 
\nonumber\\
&&+ \biggl(\frac{g'}{g}\biggr)^2 {\cal A}_{\alpha}^2 - 
(\partial_i {\cal A}_{\alpha})^2 \biggr], ~~L(\eta) 
= \int d^3 x {\cal L}(\vec{x},\eta),
\label{action1}
\end{eqnarray}
describes the evolution of the two ( $\alpha= \otimes,~\oplus$) 
transverse degrees of 
freedom defined by  the Coulomb gauge condition $A_0=0$ and 
$\vec{\nabla}\cdot\vec{A} =0$ (the prime denotes 
differentiation with respect to conformal time). The fields
$A_{\alpha}= g {\cal A}_{\alpha}$  have kinetic terms 
with canonical normalization and the time evolution 
\begin{equation}
{\cal A}_{\alpha}'' - \nabla^2 {\cal A}_{\alpha} - g ( g^{-1})'' 
{\cal A}_{\alpha} =0
\end{equation}
can be derived from the Euler-Lagrange equations
By functionally deriving the 
 the action  we get the canonically conjugated momenta 
$\pi_{\alpha} = {\cal A}'_{\alpha} + (g'/g) {\cal A}_{\alpha}$ leading 
to the Hamiltonian density and to the associated Hamiltonian
\begin{eqnarray}
&&{\cal H}(\vec{x},\eta) = \frac{1}{2}\sum_{\alpha} 
\biggl[ \pi_{\alpha}^2 + (\partial_{i} {\cal A}_{\alpha} )^2 - 
2 \frac{g'}{g} {\cal A}_{\alpha} \pi_{\alpha}\biggr],
\nonumber\\
&& H(\eta) =\int d^3 x {\cal H}(\vec{x}).
\end{eqnarray}
The operators corresponding to the classical polarizations
appearing in the Hamiltonian density 
\begin{eqnarray}
&&\hat{\cal A}_{\alpha} (\vec{x}, \eta) = \int{ d^3 k}\frac{1}{(2 \pi)^{3/2}}  
\hat{\cal A}_{\alpha}(k,\eta) e^{i \vec{k}\cdot{\vec{x}}},
\nonumber\\
&&\hat{\cal A}_{\alpha}(k,\eta) = \frac{1}{ \sqrt{ 2 k}}
\bigl[ \hat{a}_{k,\alpha}(\eta)+ \hat{a}^{\dagger}_{-k,\alpha}(\eta)\bigr],
\nonumber\\
&&\hat{\pi}_{\alpha} (\vec{x}, \eta) = \int{ d^3 k}
\frac{1}{(2 \pi)^{3/2}}  \hat{\pi}_{\alpha}(k,\eta) 
e^{i \vec{k}\cdot{\vec{x}}},
\nonumber\\
&&\hat{\cal \pi}_{\alpha}(k,\eta) 
= i \sqrt{\frac{k}{2}}\bigl[
\hat{a}_{k,\alpha}(\eta)- \hat{a}^{\dagger}_{-k,\alpha}(\eta)\bigr],
\end{eqnarray}
obey canonical commutation relations and the associated 
creation and annihilation operators satisfy
$[\hat{a}_{k,\alpha}, \hat{a}^{\dagger}_{p,\beta}] = 
\delta_{\alpha\beta}\delta^{(3)} (\vec{k} - \vec{p})$.
The Hamiltonian can then be written as:
\begin{eqnarray}
&&H(\eta) = 
\int d^3 k \sum_{\alpha}\biggl[ k~(\hat{a}^{\dagger}_{k,\alpha} 
\hat{a}_{k,\alpha}
+ \hat{a}^{\dagger}_{-k,\alpha} \hat{a}_{-k,\alpha} + 1) 
\nonumber\\
&&+ \epsilon(g) \hat{a}_{-k,\alpha} \hat{a}_{k,\alpha} + \epsilon^{\ast}(g) 
\hat{a}^{\dagger}_{k,\alpha} \hat{a}^{\dagger}_{-k,\alpha}\biggr], ~~~ 
\epsilon(g) = i \frac{g'}{g}.
\end{eqnarray}
The (two-modes) Hamiltonian 
contains  a free part and the effect of the variation 
of the coupling constant is  encoded in the 
(Hermitian) interaction term which is quadratic in the creation 
and annihilation operators whose evolution equations, read,
in the Heisenberg picture
\begin{eqnarray}
&&\frac{ d \hat{a}_{k,\alpha}}{d \eta} = - i k 
\hat{a}_{k,\alpha} - \frac{ g'}{g} 
\hat{a}^{\dagger}_{-k,\alpha},
\nonumber\\
&&
\frac{ d \hat{a}^{\dagger}_{k,\alpha}}{d \eta} = 
i k \hat{a}^{\dagger}_{k,\alpha} - \frac{ g'}{g} 
\hat{a}_{-k,\alpha}.
\label{heiseq}
\end{eqnarray}
The general solution of the previous system of equations can be written 
in terms of  a Bogoliubov-Valatin transformation
\begin{eqnarray}
&& \hat{a}_{k,\alpha}(\eta) = \mu_{k,\alpha}(\eta) \hat{b}_{k,\alpha} + 
\nu_{k,\alpha}(\eta)\hat{b}^{\dagger}_{-k,\alpha}
\nonumber\\
&& \hat{a}^{\dagger}_{k,\alpha}(\eta) 
= \mu^{\ast}_{k,\alpha}(\eta) \hat{b}^{\dagger}_{k,\alpha} + 
\nu_{k,\alpha}^{\ast}(\eta)\hat{b}_{-k,\alpha}
\label{heis}
\end{eqnarray}
where $\hat{a}_{k,\alpha}(0) = 
\hat{b}_{k,\alpha}$ and $\hat{a}_{-k,\alpha}(0) = \hat{b}_{-k,\alpha}$. 
Unitarity requires that  
 the two complex functions $\mu_{k}(\eta)$ and $\nu_{k}(\eta)$ 
are subjected to the condition $|\mu_{k}(\eta)|^2 - |\nu_{k}(\eta)|^2 =1$ 
which also implies that
$\mu_{k}(\eta)$ and $\nu_{k}(\eta)$ can be parameterized in terms of 
one real amplitude and two real phases
\begin{equation}
\mu_k = e^{i \theta_{k}} \cosh{r_{k}},
~~~~\nu_k = e^{ i(2\phi_{k} -  \theta_{k})} 
\sinh{r_{k}},
\label{sq}
\end{equation}
($r$ is sometimes called squeezing parameter and $\phi_{k}$
is the squeezing phase; from now on we will drop the subscript 
labeling each polarization if not strictly necessary).
The total number of produced photons  
\begin{equation}
\langle 0_{-k} 0_{k}| 
\hat{a}^{\dagger}_{k}(\eta) \hat{a}_{k}(\eta) + 
\hat{a}_{-k}^{\dagger} \hat{a}_{-k} |0_{k} 0_{-k}\rangle= 
2 ~\overline{n}_k.
\label{num}
\end{equation}
is expressed in terms of
 $\overline{n}_{k} =\sinh^2{r_{k}}$, i.e. the  mean number 
of produced photon pairs in the mode $k$.
Inserting  Eqs. (\ref{heis}),(\ref{sq}) and (\ref{num}) 
into Eqs. (\ref{heiseq})  
we can derive a closed system involving only the 
$\overline{n}_k$  and the related phases:
\begin{eqnarray}
&&\frac{d  \overline{n}_{k}}{d \eta} = 
-2 f(\overline{n}_{k}) \frac{g'}{g} 
\cos{2 \phi_{k}},
\label{I}\\
&& \frac{d \theta_{k}}{d\eta} = - k + \frac{g'}{g} 
\frac{\overline{n}_{k}}{f(\overline{n}_{k})}
\sin{2 \phi_{k}} ,
\label{II}\\
&& \frac{ d \phi_{k}}{ d \eta} = - k + \frac{g'}{ g} 
\frac{d f(\overline{n}_{k})}{d \overline{n}_{k}}
\sin{ 2 \phi_{k}},
\label{III}
\end{eqnarray}
where $f(\overline{n}_{k}) = \sqrt{ \overline{n}_{k}(\overline{n}_{k} + 1)}$.

In quantum optics \cite{lou} the coherence properties of light fields have 
been  a subject of intensive investigations for nearly half a century. 
Magnetic fields over galactic scales have typical frequency 
of the order $10^{-14}$--$10^{-15}$ Hz which clearly fall well outside the 
optical range. Thus, the analogy with quantum optics is only technical.
The same quantum optical analogy has been successfully exploited 
in particle \cite{4} and heavy-ions physics \cite{6}  
of pion correlations (in order to measure the size of the 
strongly interacting region) and in the phenomenological 
analysis of hadronic multiplicity distributions. 

The interference between the amplitudes of the magnetic fields (Young 
interferometry \cite{11}, in a quantum optical language) estimates 
 the first order coherence of the magnetic background at different spatial 
locations making use of the two-point correlation function  
whose trace  over the physical polarizations and for coincidental 
spatial locations is  related to the magnetic energy density.
Recall that in our gauge
\begin{equation}
\hat{B}_{k}(\vec{x},\eta) = 
\frac{i g}{(2\pi)^{3/2} a^2(\eta)}\sum_{\alpha}e^{\alpha}_{i}\int 
\frac{d^3 k}{ \sqrt{2 k}}k_{j} \epsilon_{j i k}\bigl[ 
\hat{a}_{k,\alpha}e^{i \vec{k}\cdot\vec{x}} -
\hat{a}^{\dagger}_{k,\alpha}e^{-i \vec{k}\cdot\vec{x}}\bigr].
\end{equation}
By summing up over 
the polarizations according to
\begin{equation}
{\cal K}_{ij} = 
\sum_{\alpha}  e^{\alpha}_{i}(k) e^{\alpha}_{j}(k) = \biggl( \delta_{ij} 
- \frac{k_{i} k_{j}}{k^2}\biggr), 
\end{equation}
we get that 
\begin{equation}
 {\cal G}_{ij} (\vec{r}, \eta) = \langle 0_{-k} 0_{k}|
\hat{B}_{i}(\vec{x}, \eta)
\hat{B}_{j}(\vec{x} + \vec{r},\eta) | 0_{k} 0_{-k}\rangle
\end{equation}
can be expressed, using Eqs. (\ref{heis}) and (\ref{sq}) 
\begin{equation}
{\cal G}_{ij} (\vec{r}) = \int d^3 k 
{\cal G}_{ij}(k) e^{i \vec{k}\cdot\vec{r}}.
\end{equation}
where
\begin{eqnarray}
&&{\cal G}_{ij}(k,\eta) = \frac{g^2(\eta) {\cal K}_{i j}}{2(2\pi)^3 a^4(\eta)} 
k[2\sinh^2{r_{k}} 
\nonumber\\
&&+\sinh{2 r_{k}}\cos{ 2\phi_{k}}]
\end{eqnarray}
(the vacuum contribution, occurring for $r_{k}=0$, has been 
consistently subtracted).
The intercept for $\vec{r}=0$ of the two-point function traced 
with respect to the two polarizations is related to the magnetic energy density
\begin{equation}
\frac{d \rho_{B}}{d\ln{\omega}} \simeq  
\frac{g^2(\eta)\omega^4}{2\pi^2} 
[2\sinh^2{r_{k}} + \sinh{2 r_{k}}\cos{ 2\phi_{k}}]
\end{equation}
(where $\omega= k/a$ is the physical frequency).
The two-point function and its trace only 
depend upon $\overline{n}_{k}$ and upon $\phi_{k}$. 
Since 
Eqs. (\ref{I}) and (\ref{III}) do not contain 
any dependence upon $\theta_{k}$ we can attempt to solve the time evolution
by solving them simultaneously. In terms of the new variable 
$x = k \eta$ Eqs. (\ref{I}) and (\ref{III}) can be written as 
\begin{eqnarray}
&&\frac{d \phi_{k}}{d x} = -1 + \frac{d \ln{g}}{d x} 
\frac{d f(\overline{n}_{k})}{d \overline{n}_{k}} \sin{2 \phi_{k}},
\label{IV}\\
&&\frac{d \overline{n}_{k}}{d\ln{g}} = - 2 f(\overline{n}_{k}) 
\cos{2 \phi_{k}},
\label{V}
\end{eqnarray}
If $|(d\ln{g}/dx)(d f(\overline{n}_k)/d \overline{n}_k) 
\sin{2 \phi_{k}}| > 1$, then Eqs. (\ref{IV}) and 
(\ref{V}) can be written as 
\begin{equation}
\frac{d u_{k}}{d \ln{g}} = 
2 \frac{d f(\overline{n}_{k})}{d \overline{n}_{k}} u_{k},
~~~\frac{d \overline{n}_{k}}{d \ln{g} } = 
-2 f(\overline{n}_{k}) \frac{ 1 - u_{k}^2}{1 + u_{k}^2}
\end{equation}
where $ \phi_{k}= \arctan{u_{k}}$. By trivial algebra 
we can get a differential relation between $u_k$ 
and $\overline{n}_{k}$ which can be 
exactly integrated with the result that 
$u_{k}^2 - f(\overline{n}_k) u_{k} + 1= 0$. By inverting 
this last relation we obtain two different solutions 
with equivalent physical properties, namely
\begin{equation}
u_{k}(\overline{n}_{k}) = \bigl[\frac{1}{2}(\sqrt{ 
\overline{n}_{k}(\overline{n}_{k}+ 1)} 
\pm \sqrt{\overline{n}_{k}(\overline{n}_{k} +1)  -4})\bigr].
\label{rel}
\end{equation}
If we choose the minus sign in Eq. (\ref{rel}) we obtain that 
$\phi_{k} \sim ( m + 1) \pi/2 $, $m= 0, 1,2...$
 with corrections of order 
$1/\overline{n}_k$ . In the opposite case 
 $\phi_{k} \sim \arctan{(\overline{n}_{k}/2)}$ within the same accuracy 
of the previous case(i.e. $1/\overline{n}_{k}$). 
By  using the relation between $u_{k}$ and $\overline{n}_{k}$ the condition
 $|(d\ln{g}/dx)(d f(\overline{n}_k)/d \overline{n}_k) 
\sin{2 \phi_{k}}| > 1$ is equivalent to $ x \laq 1$, 
if, as we are assuming, $|g'/g|$ vanishes 
as $\eta^{-2}$ for $\eta\rightarrow \pm \infty$ and it is, 
piece-wise, continuous.  
By inserting Eq. (\ref{rel}) into Eq. (\ref{IV})  a consistent solution 
can be obtained, in this case, if we integrate 
the system between $\eta_{f}$ and $\eta_{i}$ defined 
as the conformal times where 
$|(d\ln{g}/dx)(d f(\overline{n}_k)/d \overline{n}_k) 
\sin{2 \phi_{k}}|= 1$:
\begin{eqnarray}
&&\overline{n}_{k}(\eta_{f}) \sim 
\frac{1}{4} \biggl(\frac{g(\eta_{f})}{g(\eta_i)} 
- \frac{g(\eta_{i})}{g(\eta_{f})}\biggr)^2, 
\nonumber\\
&&\phi_{k}(\eta) \sim ( m + 1) \frac{\pi}{2}
 + {\cal O}(\frac{1}{\overline{n}_{k}(\eta)}),~~~m= 0, 1,2...
\label{xl1}
\end{eqnarray}
If $|(d\ln{g}/dx)(d f(\overline{n}_k)/d \overline{n}_k) 
\sin{2 \phi_{k}}| < 1$  (i. e. $x > 1$)  
the consistent solution of our system is given by
\begin{eqnarray}
&&\overline{n}_{k}(\eta_{f}) =
\sinh^2{\biggl(2 \int^{k\eta} \ln{g(x')} \sin{2 x'} dx'\biggr)}
\nonumber\\
&&\phi_{k} \sim - k\eta + \varphi_{k}, ~~~\varphi_{k} \simeq {\rm constant}.
\end{eqnarray}
If the coupling constant evolves 
continuously between $-\infty$ and $+\infty$ with a (global) maximum located 
at some time $\eta_r$ then, for $x> 1$,
 $\overline{n}_{k} \sim {\rm const.}$. Indeed 
by taking trial functions with bell-like shape for $|g'/g|$ we can show
that $\overline{n}_{k}$ oscillates around zero for large $\phi_{k}$.

Up to now our considerations were general. Let us give some 
specific example of our technique.
In the low energy phase ($\eta<\eta_{s}$)  of the 
pre-big-bang evolution the dilaton coupling is determined by 
the variation of the low-energy effective action \cite{ven}:
\begin{equation}
a(\eta) \simeq |\eta|^{-\frac{1}{\sqrt{3}+ 1}},~~ g(\eta) \simeq 
|\eta|^{-\sqrt{3}/2} ~~\eta<\eta_{s}.
\label{a}
\end{equation}
During the stringy phase the average time evolution of the
coupling constant can  be described by:
\begin{equation}
a(\eta)\simeq \eta^{-1},~~
g(\eta) \simeq  |\eta|^{-\beta} 
,~~\eta_{s}<\eta<\eta_{r},
\label{b}
\end{equation}
where $\beta = - (\phi_s -
\phi_r)/(2\ln{z_{s}})$ where $z_{s} = \eta_{s}/\eta_r$.
For  $\eta>\eta_r$, the background is dominated by radiation
(i.e. $a(\eta)\simeq\eta$) and the coupling 
constant freezes to its constant value 
(i.e. $\phi=\phi_{r}={\rm const.}$ for $\eta>\eta_{r}$). 
Notice that $g(\eta_{r})= \exp{(\phi_{r}/2)} = g_{r} \simeq 0.1$--$0.01$.
For $k\eta < 1$ we have, from Eqs. (\ref{xl1}) that
\begin{equation}
\overline{n}_{k}(\eta_{r})\simeq  |\eta_i/\eta_{r}|^{2\beta} \sim 
|k/k_{r}|^{-2\beta},~~~k_{s}<k<k_{r}
\label{mean1}
\end{equation}
where $|\eta_{i} |\sim k^{-1} <|\eta_{s}|$. Similarly, 
if $ |\eta_{i}| >|\eta_{s}|$,  
\begin{equation}
\overline{n}_{k}(\eta_r) \simeq  |k/k_{s}|^{-\sqrt{3}} 
|g(\eta_{s})/g(\eta_{r})|^{-2}
,~~~k<k_{s}~~~.
\label{mean2}
\end{equation}

Notice that we assumed $\beta >0$ which means that the coupling constant 
does not decreases during the stringy phase.
Due to magnetic flux conservation \cite{kra,zel}
the fraction of electromagnetic energy stored in the
mode $\omega$ does not change and it is defined as 
\begin{eqnarray}
&&\lambda(\omega) = \frac{1}{\rho_{\gamma}}\frac{d\rho_{B}}{d\ln{\omega}} =
\frac{g^2}{4\pi}\biggl(\frac{\omega}{\omega_{r}}\biggr)^4
\overline{n}_{k}(\eta_{r}) \sin^2{k\eta},
\label{lam}\\
&&\rho_{\gamma}(\eta) =M_{P}^2
H_{r}^2 \left(\frac{a_r}{a}\right)^4\equiv 
\omega_{r}^4 \left({\frac{g_r}{4\pi}}\right)^2
\end{eqnarray}
where $\omega_{r}\sim
a_{r}/\eta_{r}=\sqrt{g_{r}/4\pi}10^{11}$ Hz is the maximal amplified
frequency red-shifted today and where 
we assumed $\overline{n}_{k}(\eta_{r})>1$. Notice that 
in the unifying notation of eq. (\ref{lam}) the oscillating part
occurs, for each mode, when $k\eta>1$ but not in the opposite limit 
where $\phi_{k}\sim \pi/2$. 
The critical density constraint, implies, 
during the stringy phase that $\beta< 2$.
If $\beta \laq 2$ (for instance $\beta\simeq 1.9$) we can have that  
$\lambda(\omega_{\rm dec} )= g_r^2 
(\omega_{\rm dec}/\omega_{r})^{4 - 2 \beta} \sim 
10^{-8}$ for $\omega_{\rm dec} \sim 10^{-16}$ Hz (for $\omega_{\rm dec} 
>\omega_s$). Recall that, in order to rotate the 
polarization plane of the Cosmic Microwave Background Radiation, we need, at 
decoupling, $B \gaq 10^{-3}$ Gauss, or in our language, 
$\lambda(\omega_{\rm dec}) \gaq 10^{-8}$ \cite{rot}. Similarly, at the 
scale of $1$ Mpc (i.e. $\omega_{G} \sim 10^{-14}$ Hz) we can have 
$\lambda(\omega_{G}) > 10^{-10}$ \cite{1}. Both at the galactic 
and decoupling frequencies $\overline{n}_{\omega} \gg 1$ in the framework 
of this  specific model and the quantum mechanical state is strongly 
squeezed ($|r_{k}| \gg 1$).

The quantum degree of second order coherence is 
a measure of the correlation of
the magnetic field intensities at two space-time 
points. 
The intensity fluctuations of a 
given light field are described by the 
Glauber correlation function \cite{lou}
which is nothing but the quantum mechanical 
generalization of the correlation between 
the classical intensities of two light beams. 
In quantum optics one deals mainly 
with intensity correlations of electric fields.
This is due to the fact that the photons 
of the visible part of the electromagnetic spectrum 
are detected via photo-electric effect, and, 
therefore, what is indeed detected is an electric 
current induced by a photon. In our case 
we are mainly interested in the intensity 
correlations of the magnetic part of the field and 
we can write the corresponding correlation  
$\Gamma(\vec{r})$
of the intensity operators as 
\begin{equation}
\frac{\langle : \hat{\beta}^{-}(\vec{x}, \eta) ~
\hat{\beta}^{-}(\vec{x} + \vec{r}, \eta )~
\hat{\beta}^{+} (\vec{x} + \vec{r}, \eta)~
\hat{\beta}^{+}(\vec{x}, \eta):\rangle}{
\langle : \hat{\beta}^{-}(\vec{x},\eta) \hat{\beta}^{+}(\vec{x},\eta):\rangle
\langle : \hat{\beta}^{-}(\vec{x}+\vec{r},\eta): \rangle}.
\label{glauber}
\end{equation} 
In this (normal ordered) definition the field operators refer to a single 
polarization of the field, namely
\begin{equation}
\hat{\beta}^{+}(\vec{x},\eta) = \frac{i g}{(2\pi)^{3/2}}
 \int d^3 k \sqrt{\frac{k}{2}} 
\hat{a}_{k}(\eta) e^{i \vec{k}\cdot\vec{x}}, 
\end{equation}
with $\hat{\beta}^{-}=({\hat{\beta}^{+})^{\dagger}}$.
$\Gamma(\vec{r})$ describes correlation between 
intensities in the case where the photons are detected 
simultaneously in time but at different spatial locations. 
The statistical properties of the given 
quantum mechanical state of the field 
are encoded in the intercept of the Glauber 
function namely $\Gamma(0)$.

By using Eqs. (\ref{heis}) into Eq. (\ref{glauber}) we obtain 
\begin{eqnarray}
&&\Gamma(\vec{r})= \frac{ \int d^3 k ~k |\nu_{k}(\eta)|^2 
\int d^{3} p ~p |\nu_{p}(\eta)|^2 }
{\int d^3 k~k |\nu_{k}(\eta)|^2 \int  d^3 p ~p |\mu_{p}(\eta)|^2} 
\nonumber\\
&+&\frac{\int d^3 k ~k |\nu_{k}(\eta)|^2
\int d^3 p~ p |\nu_{p}(\eta)|^2 e^{i (\vec{k} - \vec{p}) \cdot\vec{x}}}
{\int d^3 k~k |\nu_{k}(\eta)|^2 \int  d^3 p ~p |\mu_{p}(\eta)|^2}
\nonumber\\
&+&\frac{\int d^3 k~ \int d^3 p~p ~ k~ \nu_{k}^{\ast}(\eta)\mu_{k}^{\ast}(\eta)
\nu_{p}(\eta) \mu_{p}(\eta)e^{i (\vec{k}- \vec{p})\cdot\vec{r}}}
{\int d^3 k~k |\nu_{k}(\eta)|^2 \int  d^3 p ~p |\mu_{p}(\eta)|^2}
\end{eqnarray}
which in the limit $|\vec{r}| \rightarrow 0$ becomes 
\begin{equation}
\Gamma(0) = 2 + \frac{\int k~d^3 k \int p~d^3 p  \nu^{\ast}_{k}(\eta) 
\mu^{\ast}_{k}(\eta)\nu_{p}(\eta) \mu_{p}(\eta) }{\int d^3 k~k 
|\nu_{k}(\eta)|^2 \int d^3 p~p |\nu_{p}(\eta)|^2}.
\end{equation}
In order to interpret this formula we can introduce a 
further simplification, namely we can restrict our attention 
to a single mode of the field. Then, $\mu_{k}(\eta) \rightarrow \mu_{K}(\eta)
\delta^{(3)}(\vec{k} - \vec{K})$ and  $\nu_{k}(\eta) 
\rightarrow \nu_{K}(\eta) \delta^{(3)}(\vec{k} - \vec{K})$.
Thus $\Gamma(0) = 3 + |\overline{\nu}_{K}|^{-2}$.  As 
we showed, large variations in $g(\eta)$ give   $|\nu_{K}|^2 \gg 1$ .
In the case of a coherent state the intercept of the 
Glauber function is exactly one, namely $\Gamma(0) =1$ \cite{lou}. This 
property is in direct correspondence with the Poissonian 
character of the statistics. In the case of a thermal state 
(i.e. ``white light''), $\Gamma(0) \rightarrow 2$ \cite{lou}.
In the case of squeezed relic photons 
for large number of particles in each Fourier mode
$\Gamma(0) \rightarrow 3$. Since $\Gamma(0)$ represents the 
probability of two photons arriving at the same location, this is referred to 
as photon bunching. Conversely, a field with sub-Poissonnian
statistics will have $\Gamma(0)<1$, an effect known as photon anti-bunching 
in the context of the Hanbury Brown-Twiss interferometry \cite{5}.

\end{document}